\documentclass[10pt] {article} %

\usepackage[T1]{fontenc} %
\usepackage[normalem]{ulem} %

\usepackage[usenames,dvipsnames,svgnames,x11names]{xcolor}

\usepackage{cite}

\usepackage{balance}

\usepackage{listings}

\lstdefinelanguage{XML}
{
basicstyle=\ttfamily\footnotesize,
  morestring=[b]",
  moredelim=[s][\bfseries\color{Maroon}]{<}{\ },
  moredelim=[s][\bfseries\color{Maroon}]{</}{>},
  moredelim=[l][\bfseries\color{Maroon}]{/>},
  moredelim=[l][\bfseries\color{Maroon}]{>},
  morecomment=[s]{<?}{?>},
  morecomment=[s]{<!--}{-->},
  commentstyle=\color{gray},
  stringstyle=\color{blue},
  identifierstyle=\color{red}
}

\usepackage{moreverb}

\usepackage[nounderscore]{syntax}

\usepackage[pdftex]{graphicx}
\graphicspath{{./figures/}}
\DeclareGraphicsExtensions{.pdf}

\usepackage[cmex10]{amsmath}
\usepackage{amssymb}
\usepackage{mathtools}
\usepackage{amsthm}
\usepackage{amsfonts}
\usepackage{gensymb}

\usepackage{subfig} %

\usepackage{algorithmicx}
\usepackage{algpseudocode}
\usepackage[ruled]{algorithm}
\definecolor{light-gray}{gray}{0.75}
\algrenewcommand{\algorithmiccomment}[1]{\hskip3em{{\footnotesize \textcolor{light-gray}{$\blacktriangleright$}}} #1}

\usepackage{multirow} %
\usepackage{rotating} %
\usepackage{booktabs} %
\usepackage{colortbl} %
\usepackage{tablefootnote} %

\usepackage{array}
\newcolumntype{L}[1]{>{\raggedright\let\newline\\\arraybackslash\hspace{0pt}}m{#1}}
\newcolumntype{C}[1]{>{\centering\let\newline\\\arraybackslash\hspace{0pt}}m{#1}}
\newcolumntype{R}[1]{>{\raggedleft\let\newline\\\arraybackslash\hspace{0pt}}m{#1}}

\usepackage[pdftex,colorlinks=true,urlcolor=blue,citecolor=blue]{hyperref}

\usepackage{xspace}

\usepackage{enumitem}

\usepackage{blindtext}

\newcommand{\rp}{\textsc{Ripple}\xspace}

\begin{document}

\title{\rp: Scalable Incremental GNN Inferencing on Large Streaming Graphs
\thanks{~Preprint of paper to appear in the proceedings of the 45th IEEE International Conference on Distributed Computing Systems (ICDCS): Pranjal Naman and Yogesh Simmhan, “\rp: Scalable Incremental GNN Inferencing on Large Streaming Graphs,” in \textit{IEEE International Conference on Distributed Computing Systems (ICDCS)}, 2025}
}

\author{Pranjal Naman and Yogesh Simmhan\\
Department of Computational and Data Sciences (CDS),\\
Indian Institute of Science (IISc),\\
Bangalore 560012 India\\
Email: \{pranjalnaman, simmhan\}@iisc.ac.in
}

\date{}
\maketitle

\begin{abstract}
Most real-world graphs are dynamic in nature, with continuous and rapid updates to the graph topology, and vertex and edge properties. 
Such frequent updates pose significant challenges for inferencing over Graph Neural Networks~(GNNs). Current approaches that perform \textit{vertex-wise} and \textit{layer-wise} inferencing are impractical for dynamic graphs as they cause redundant computations, expand to large neighborhoods, and incur high communication costs for distributed setups, resulting in slow update propagation that often exceeds real-time latency requirements. 
This motivates the need for streaming GNN inference frameworks that are efficient and accurate over large, dynamic graphs.
We propose \rp, a framework that performs fast incremental updates of embeddings arising due to updates to the graph topology or vertex features. \rp provides a generalized incremental programming model, leveraging the properties of the underlying aggregation functions employed by GNNs to efficiently propagate updates to the affected neighborhood and compute the exact new embeddings. 
Besides a single-machine design, we also extend this execution model to distributed inferencing, to support large graphs that do not fit in a single machine's memory. \rp on a single machine achieves up to $\approx28000$~updates/sec for sparse graphs like Arxiv and $\approx1200$~updates/sec for larger and denser graphs like Products, with latencies of $0.1$ms--$1$s that are required for near-realtime applications. The distributed version of \rp offers up to $\approx30\times$ better throughput over the baselines, due to $70\times$ lower communication costs during updates.
\end{abstract}

\section{Introduction}
\label{sec: intro}
Graph Neural Networks~(GNNs) have the ability to learn low-dimensional representations that capture both the topology and the attributes of graph datasets. This makes them a powerful tool for wide-ranging linked data applications, such as detecting financial frauds~\cite{liu2021pick, dou2020enhancing}, predicting traffic flow~\cite{guo2019traffic}, analyzing social networks~\cite{yang2021gnnsocial, kipf2016semisupervised} and making eCommerce recommendations~\cite{chen2020tgcn, chang2021sequential, wu2022gnnrec, lyu2020gnnrec2}.
These applications often handle \textit{large graphs}, with millions--billions of vertices and edges, e.g., millions of users performing transactions with a fintech app or 100,000s of traffic junctions in a city with traffic flow details. They also impose \textit{real-time latency} requirements during GNN inferencing (or ``serving''~\cite{wu2023inkstream}). 
For instance, a delay in identifying suspicious transactions in financial networks can allow fraudulent activities to occur, leading to financial loss. Similarly, delays in real-time traffic flow prediction used for traffic signal optimization can exacerbate congestion during peak hours. 
Additionally, most applications also require \textit{deterministic} inference responses. E.g., the same transaction should be classified as a fraud across multiple inferencing requests, in the absence of any other changes.

\begin{figure}[t]
    \centering
    \includegraphics[width=0.8\columnwidth]{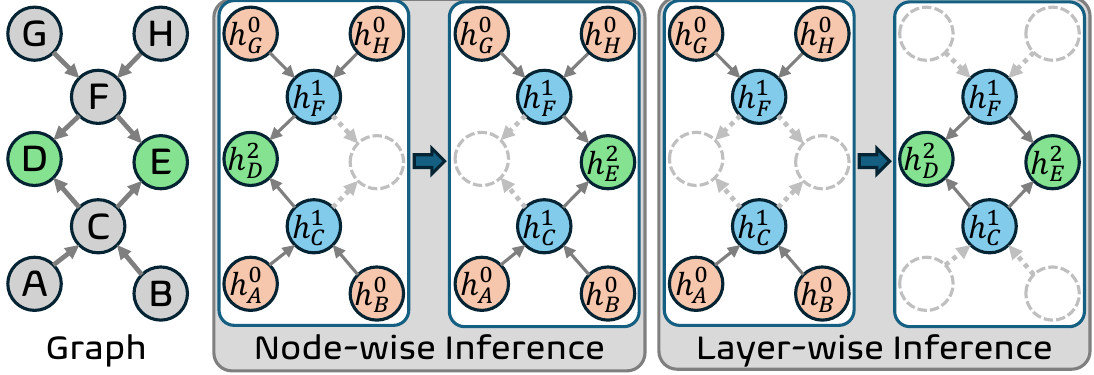}
    \caption{\textit{Vertex-wise} vs. \textit{Layer-wise} inferencing on static graphs.}
    \label{fig:static-inference}
\end{figure}

\begin{figure}[t]
    \centering
    \subfloat[Effect of neighborhood sampling on vertex-wise inference accuracy and latency on Reddit graph.\label{subfig: inf-fanout-spread}]{\includegraphics[width=.42\linewidth]{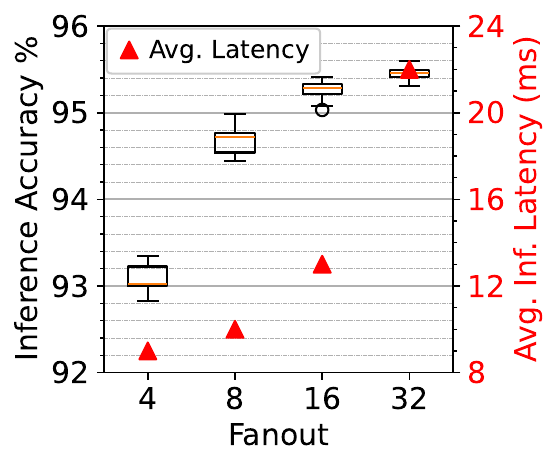}}%
    ~
    \subfloat[\% of affected vertices and inference latency per batch-update, for proposed \rp and baseline recomputing (RC) models, with differing update batch sizes for two graphs.\label{subfig: affected-nodes}]{\includegraphics[width=.575\linewidth]{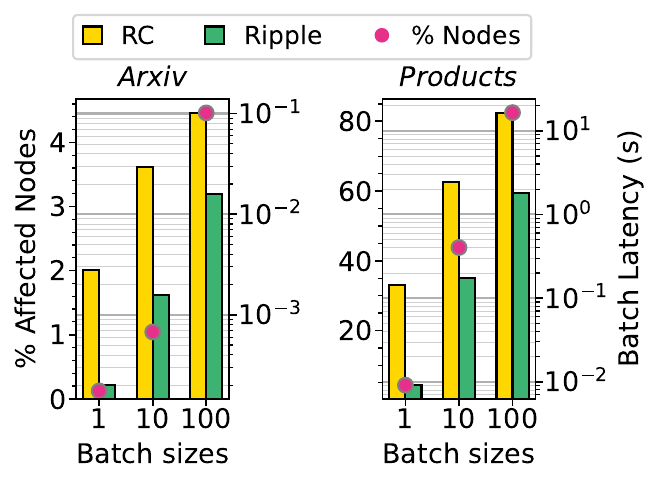}}
    \caption{Performance and accuracy trade-offs for vertex-wise and layer-wise inferencing using $3$-Layer SAGEConv model.}
    \label{fig: motivation-plot-1}
\end{figure}

Given these low-latency needs, performing this GNN inferencing offline once, on all graph entities, and returning the pre-computed result rapidly using a simple lookup is an efficient approach to serving~\cite{yin2022dgi}.
There are two ways to do this. In \textit{vertex-wise inferencing} (Fig.~\ref{fig:static-inference}, center), the embeddings within an $L$-hop neighborhood of each target vertex (green) being labeled are aggregated, assuming a vertex labeling GNN task. This is akin to the forward pass of GNN training. While training, the neighborhood of the computational graph is randomly sampled to keep the subgraph size manageable and still achieve good model accuracy~\cite{hamilton2017sage}. However, such sampling affects the correctness and deterministic nature of the predictions during inference~\cite{kaler2022accelerating}. Hence, the entire $ L$-hop neighborhood is considered when inferencing, leading to \textit{neighborhood explosion}, with higher memory and computational demands. Fig.~\ref{subfig: inf-fanout-spread} shows this trade-off for the \textit{Reddit} social network graph~($233K$ vertices, $114M$ edges; experiment setup described in \S~\ref{subsec: eval-expsetup}). As the sampling fanout size increases, we get a better/deterministic accuracy~(left Y axis) but also a longer per-vertex average inferencing time~(right Y axis). 

So, a \textit{layer-wise} inferencing approach (Fig.~\ref{fig:static-inference}, right) is typically employed for bulk processing~\cite{yin2022dgi}. Here, the embeddings for each hop~(colored vertex layers) are computed for all vertices in the graph and used as inputs to calculate the embeddings for the next layer (orange to blue to green).
This avoids large computational graphs and prevents redundant computation due to overlaps in computation graphs of proximate vertices, which occurs in vertex-wise inferencing,
e.g., $h^1_F$ and $h^1_C$ are recomputed for inferencing of $D$ and $E$.
As we show later~(Fig.~\ref{fig: gpu-analysis}), layer-wise computation~(DRC) is $\approx2\times$ faster than vertex-wise computation~(DNC) and is used in this paper as the basic inferencing approach.

Adding to the complexity, applications using GNN inferencing often operate on graphs that \textit{evolve continuously} in the presence of vertex/edge additions/deletions or changes to their features. E.g., a friend being added or removed in a social network causing an edge addition/deletion~\cite{bai2020temporal}, changing traffic flows in a road network that updates an edge feature~\cite{guo2019traffic}, or transactions in a fintech network that change the account-balance vertex feature~\cite{song2023financial}. Thousands of such dynamic graph updates can occur per second, affecting the output of GNN predictions for entities in their neighborhood~\cite{wu2020deltagrad, mahadevan2023cost}.

However, performing layer-wise inference on the \textit{whole graph} every time an update (or a small batch of updates) occurs is too time-consuming for latency-sensitive applications, e.g., taking $\approx20$~seconds for updating the predicted labels for vertices of the \textit{Products} graph~($2.5M$ vertices, $123M$ edges).
Instead, we can limit the recomputation of the vertex labels only to those that fall within the $L$-hop neighborhood of entities that have been updated in the graph.
Here, the cascading effects of the updates on the other entities affected in the graph grow with the average in-degree of the graph and the number of updates accumulated in a batch. E.g., in Fig.~\ref{subfig: affected-nodes}, the fraction of vertices affected for the \textit{Arxiv} graph (in-degree $\approx7$) increases from $0.1\%$ to $\approx4\%$ of the total vertices (red markers on left Y axis), as the update batch sizes increase from $1$ to $100$. However, for the \textit{Products} graph~(in-degree $\approx50$), the growth is sharper, from $4\%$ to $\approx80\%$.

This also directly impacts the \textit{latency} to translate incoming graph updates into updated GNN embeddings and inversely affects the \textit{update throughput} that can be supported.
As the affected neighborhood size increases, so does the per-batch latency to perform \textit{inference recomputations (RC)} upon updates, growing from $3$ms to $100$ms for \textit{Arxiv} using batch sizes from 1--100, and from $150$ms to $17$s for \textit{Products}~(Fig.~\ref{subfig: affected-nodes}, RC yellow bars on right Y axis). The throughput of updates processed to give fresh GNN predictions is between $180$--$1000$up/s for Arxiv and $3$--$7$up/s for Products.
This still leaves room for improvement when higher update rates need to be supported for larger and denser graphs, such as \textit{Products}. 

One key intuition that we leverage in this paper is to \textit{avoid redundant computation} of embeddings for the entire neighborhood of updates and instead scope it to just a subset of the operations. 
In particular, when one~(or a few) vertices ($V'$) among the in-neighbors ($V$) of another vertex ($u$) get updates, rather than aggregate the embeddings from all $V$ vertices, we \textit{incrementally aggregate} only the deltas from the updated vertices $V'$. This reduces the number of computations from $k=|V|$ to $k'=|V'|$, which is often an order of magnitude smaller.

The layer-wise GNN inferencing operations require the whole graph and its embeddings to be loaded and present in memory. Interestingly, layer-wise GNN inferencing does not benefit much from using GPUs since the computational load is modest~(Fig.~\ref{fig: gpu-analysis}). Instead, CPUs offer comparable performance and also benefit from larger memory. But graphs such as \textit{Papers}~($111M$ vertices, $1.6B$ edges, $128$ features)~\cite{hu2020open} and their embeddings can easily take $>128$~GiB of RAM.
This necessitates efficient \textit{distributed execution} of incremental computation on a compute cluster to support larger graphs. 

In this paper, we present \rp, a low-latency framework for streaming GNN inference that efficiently handles real-time updates to large-scale graphs using incremental computation. It uniquely 
applies a delta to undo previous embedding aggregations, and redoes them using updated embeddings to sharply reduce any recomputation. \rp offers both a single-machine and a distributed execution model.

Specifically, we make the following contributions:
\begin{enumerate}[leftmargin=*]
    \item  We propose \rp, an incremental \textit{layer-wise} GNN inferencing framework that supports streaming edge additions/deletions and vertex feature changes, for GNN models based on linear aggregation functions~(\S~\ref{sec:design}).
    \item We extend \rp to distributed incremental GNN inferencing, to support large graphs exceeding a single machine's memory (\S~\ref{sec:design:distr}).
    \item Our detailed comparative experiments with State-of-the-Art (SOTA) baselines for large graphs show \rp to outperform on both throughput and latency, in both single machine (up to $\approx28000$~up/s) and distributed settings (up to $\approx30\times$ better throughput) (\S~\ref{sec:evaluation}).
\end{enumerate}
We also provide background on GNN training/inference (\S~\ref{sec: background}), contrast \rp with related works (\S~\ref{sec:related}), and conclude with a discussion of future directions~(\S~\ref{sec: conclusion}).

\section{Background}
\label{sec: background}

\subsection{Training vs Inference of Graph Neural Networks}
\label{subsec: background-train-v-inf}

GNNs are trained using a forward and backward pass, similar to Deep Neural Networks (DNNs).
During the \textit{forward pass}, when training over a labeled vertex $u$, the layer-\textit{l} of an \textit{L}-layer GNN uses the \textsc{Aggregate} function to accumulate the $l - 1$ embeddings of the neighbors $\mathcal{N}(u)$ of the vertex. The aggregated embedding vector $x^l_u$ is then processed by an \textsc{Update} function, which is the learnable component of the network, and then passed through a nonlinear function $\sigma$. This is repeated $L$-times to get the final layer embedding $h^L$ for each labeled training vertex, marking the end of a forward pass. This is followed by a \textit{backward pass} to update the learnable parameters based on the loss function. 
\begin{align}
    x^l_u =&~ \textsc{Aggregate}^l(\{h^{l-1}_v, v \in \mathcal{N}(u)\}) \label{eq:gnn}\\
    h^l_u =&~ \sigma(\textsc{Update}^l(h^{l-1}_u, x^l_u))
\end{align}

During GNN inference, only the forward pass is necessary as the parameters have already been learned and remain fixed. For static graphs, 
the final layer embeddings and corresponding labels for all vertices in the graph can be pre-calculated, 
thus avoiding computation at inference time.
As discussed, a \textit{layer-wise} approach~\cite{yin2022dgi} avoids redundant recomputation and \textit{neighborhood explosion} evident in \textit{vertex-wise} inference approach.

\subsection{Inference on Streaming Graphs}
\label{subsec:bg-inf}
\begin{figure}[t]
    \centering
    \includegraphics[width=0.77\linewidth,trim={0cm 6cm 0 4cm},clip]{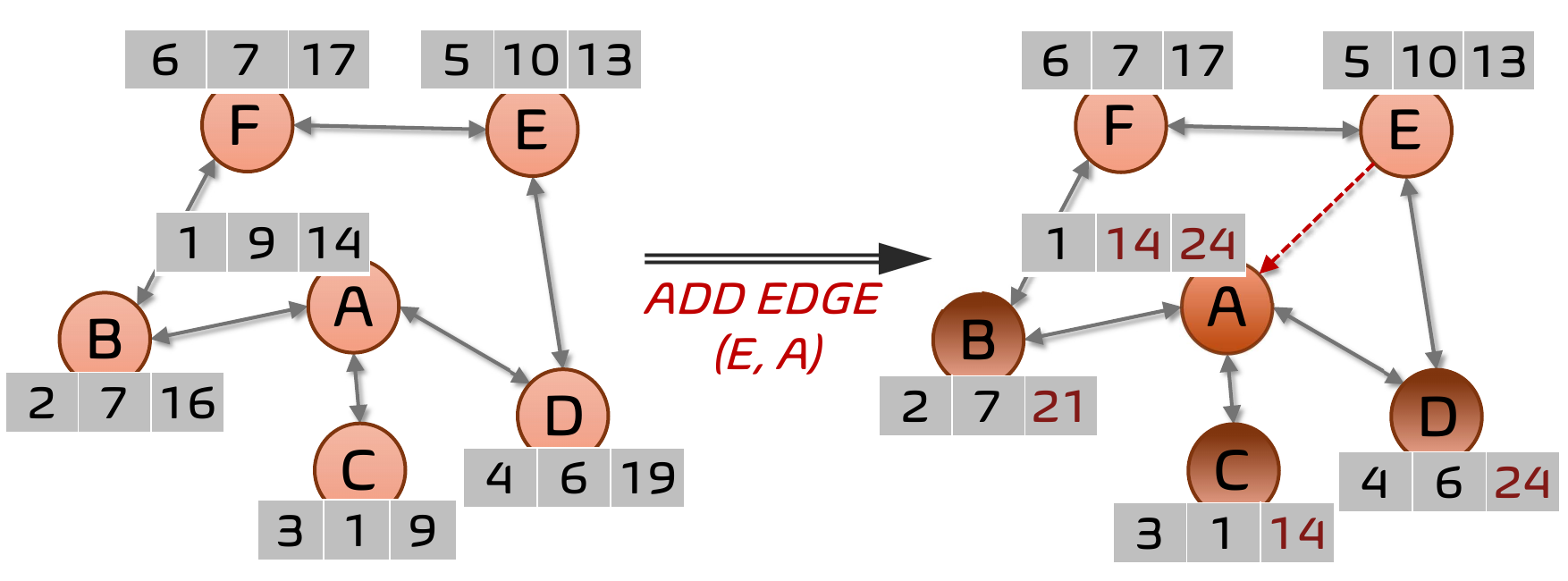}
    \caption{Cascading effect on the vertex embeddings due to an edge addition during GNN inference.
    }
    \label{fig:streaming-edge-add}
\end{figure}

Unlike for static graphs, pre-calculating the final layer embeddings and labels for all vertices is not beneficial for dynamic graphs. Each update to a vertex or edge triggers a cascade of embedding updates across multiple vertices at each hop, ultimately changing the embeddings and, consequently, the predicted labels of all vertices at the final hop. 

Fig.~\ref{fig:streaming-edge-add} shows a unit-weighted graph with pre-calculated embeddings for a $2$-layer GNN with $sum$ as the aggregator. When edge $(E, A)$ is added, the $h^1$ and $h^2$ embeddings of $A$ get updated, which cause cascading updates to the $h^2$ embeddings of $\{B, C, D\}$, possibly changing the labels of $\{A, B, C, D\}$ in the final hop. Notably, the embeddings of $F$ and $E$ remain unaffected. Any change in the graph topology or the vertex features only ``ripples'' through to a maximum of $L$-hops, for an $L$-layer GNN, making it unnecessary to update the embeddings of all vertices. 
The fraction of affected vertices can span from a few percent of vertices for \textit{Arxiv} to a large fraction of vertices for \textit{Products} (Fig.~\ref{subfig: affected-nodes}). 
It is crucial to propagate the effect of the dynamic graph updates quickly and accurately to ensure timely inference.

We can classify the inference of streaming graphs into \textit{trigger-based} and \textit{request-based}. \textit{Trigger-based} inferencing notifies the application of changes to the predicted vertex label due to graph updates immediately. In contrast, \textit{request-based} inference follows a pull-based model, where the application queries the label of a specific entity in the graph as needed. Both classes have unique requirements and different strategies,
e.g., update propagation can be done lazily for rarely-accessed vertices in the graph for \textit{request-based} inference model. In this paper, we tackle \textit{trigger-based} inferencing.

\subsection{Aggregation Functions}
In a multi-layered GNN, the \textit{neighborhood aggregation} forms a key operation, where each vertex combines the information from its neighbors to generate its representation using Eqn.~\ref{eq:gnn}. Existing research on GNNs often uses \textit{linear} functions like $sum$, $mean$, or $weighted~sum$ based on edge weights; occasionally, they consider \textit{non-linear} aggregation functions like $max$ or $min$. However,
\textit{linear aggregation functions} are more popular, e.g., the Graph Convolution Networks~(GCN) architecture~\cite{kipf2016semisupervised} adopts a weighted summation while GraphSAGE~\cite{hamilton2017sage} typically uses sum or mean. Graph Attention Networks~(GAT)~\cite{velivckovic2017graph} follow an attention-based summation where the attention on an edge is calculated based on the features of the participating vertices. Xu et al.~\cite{xu2018powerful} show that the expressiveness of the sum aggregator overshadows other functions, and use this in Graph Isomorphism Networks~(GIN). Dehmamy et al.~\cite{dehmamy2019understanding} and Corso et al.~\cite{corso2020principal} show that the expressive power of a GNN module can be increased by using a combination of aggregators, e.g., the former uses a combination of sum and mean, while the latter uses both linear and non-linear aggregators. Given the popularity of linear functions, we focus on such GNN architectures (Table~\ref{tab:gnn_aggregation}).
 
\begin{table}
\centering
\caption{Popular Linear Aggregation functions in GNNs.}
\label{tab:gnn_aggregation}
\begin{tabular}{l|l|l}
\hline
\textbf{Aggregation Fn.} & \textbf{Description} & \textbf{Used By} \\ \hline\hline
$sum$            & \(\mathbf{h}_i = \sum_{j \in \mathcal{N}(i)} \mathbf{h}_j\) & \cite{hamilton2017sage}\cite{xu2018powerful}\cite{dehmamy2019understanding}\cite{velivckovic2017graph} \\ \hline
$mean$           & \(\mathbf{h}_i = \frac{1}{|\mathcal{N}(i)|} \sum_{j \in \mathcal{N}(i)} \mathbf{h}_j\) & \cite{hamilton2017sage}\cite{dehmamy2019understanding} \\ \hline
$weighted~sum$   & \(\mathbf{h}_i = \sum_{j \in \mathcal{N}(i)} \alpha_{ij} \cdot \mathbf{h}_j\)& \cite{kipf2016semisupervised} \\ \hline
\end{tabular}
\end{table}

\section{Related Work}
\label{sec:related}

GNNs have found numerous applications due to their ability to learn low-level representations of graph data. E-commerce platforms like Alibaba leverage GNNs to analyze user behavior and deliver personalized product recommendations~\cite{zhu2019aligraph}. Social media platforms like Pinterest utilize GNNs for content recommendation to users~\cite{ying2018pinterest}. 
Google applies GNNs in estimating travel times for Google Maps~\cite{derrow2021google}. 
Here, we focus on GNN applications over large dynamic graphs with millions--billions of vertices and edges, receiving streaming updates on the order of 100--1000s of updates per second, as seen commonly in social, eCommerce, fintech, and road networks.

Research on GNN inferencing and serving has largely been limited to static graphs~\cite{yin2022dgi,zhou2021accelerating}, where the graph structure and vertex/edge features do not change over time. Popular frameworks like DGL~\cite{wang2019deep} and Pytorch Geometric~\cite{fey2019fast} are primarily designed for GNN training and are not optimized for GNN serving. 
DGI~\cite{yin2022dgi} proposes a \textit{layer-wise} inference approach, which computes embeddings layer-by-layer and handles the tasks of all target vertices in the same layer batch-by-batch. This avoids the neighborhood explosion problem encountered while adopting \textit{vertex-wise} inference.
Frameworks such as GNNIE~\cite{mondal2022gnnie} and GraphAGILE~\cite{zhang2023graphagile} introduce specialized hardware accelerators designed to enable low-latency GNN inference. 
Hu et al.~\cite{hu2024lgrapher} propose $\lambda$-Grapher, which is a serverless system for GNN serving that achieves resource efficiency through computation graph sharing and fine-grained resource allocation for the separate memory and compute-bound requirements of a GNN layer. These methods, however, are non-trivial to extend to dynamic graphs.

Real-world graphs are often dynamic in nature and require specialized frameworks. \textit{Streaming graph analytics} has been extensively studied in literature for traditional algorithms like PageRank and ShortestPath~\cite{mariappan2021dzig,vora2017kickstarter}, and even incremental graph processing frameworks exist~\cite{taris}. Applications like PageRank on streaming graphs negate the influence of a vertex's previous state before propagating its updated state in each iteration. However, traditional solutions for streaming graph analytics are not directly applicable to GNN inferencing~\cite{wu2023inkstream} since GNNs handle large hidden states, making computation and memory footprint more demanding. Also, graph updates can extend to $L$ hops per update rather than be limited to immediate neighbors. This necessitates specialized approaches to streaming graph GNN inferencing.

There is preliminary work on GNN processing for streaming graphs.
DGNN~\cite{yao2020dygnn} proposes an LSTM-inspired architecture to capture the dynamic nature of evolving graphs, but is limited to GNN training. 
Omega~\cite{kim2025omega} proposes selective recomputation of embeddings using a heuristic to minimize the approximation errors due to stale precomputed embeddings. They also offer parallelism strategies for graph calculations to balance inference latency and accuracy. STAG~\cite{wang2023stag} solves the \textit{request-based} inferencing problem using a collaborative serving mechanism that balances the inference and staleness latencies. In contrast, \rp does \textit{trigger-based} inferencing while performing exact~(not approximate) embeddings update.

InkStream~\cite{wu2023inkstream} also addresses \textit{trigger-based} inferencing but for \textit{monotonic aggregation functions} like $min$ and $max$. They incrementally update the vertex embeddings as the graph evolves, identifying \textit{resilient} vertices whose embeddings do not change upon an update. They prune the branches in the propagation tree originating from such vertices, thus reducing the need for a full recomputation of the entire network.
However, it is confined to monotonic aggregation functions and does not support the popular class of accumulative linear functions, which are more challenging to propagate. A vertex can remain unaffected despite a change in some of its inputs due to the properties of $min$/$max$. So, the updates can propagate faster due to the pruning of the propagation tree. In contrast, linear aggregation functions like $sum$ are accumulative, and any change in even one of the inputs leads to a change in the vertices' embeddings, thus making the update propagation more intensive. Our work efficiently supports such functions while ensuring exact incremental computation of embeddings.

Further, none of these methods evaluate their solutions on large graphs with millions of vertices and $100M+$ edges. So their scalability on single machines or in a distributed setting remains unknown.
The largest public graph for GNNs, Papers~\cite{hu2020open}, is itself modest in size compared to web-scale graphs that can exceed the RAM of a single machine when considering their features and embeddings.
This highlights the scalability challenges of such methods 
for GNN inferencing over large-scale real-world graphs.

\section{\rp System Design}
\label{sec:design}

\begin{table}[t]
\centering
\caption{Summary of notations used}
\label{tab: notations}
\begin{tabular}{c||l}
\hline
\textbf{Notation}    & \textbf{Description}                                      \\ \hline \hline
$G (V, E)$            & Graph $G$ with set $V$ of vertices and set $E$ of edges                 \\ \hline
$n, m$                & $n = |V|$ number of vertices, $m = |E|$ number of edges                                   \\ \hline
$X$                   & Vertex features. Same as $H^0$.                          \\ \hline
$H^l$                 & Layer $l$ embeddings for all vertices.                    \\ \hline
$W^l$                 & Layer $l$ weights of the trained GNN.                  \\ \hline
$h_u^l$               & Embedding of vertex $u \in V$ for layer $l$                         \\ \hline
$\mathcal{N}_l(u)$    & Set of vertices at distance $l$ from $u \in V$, $\{v \mid d(u,v)=l\}$      \\ \hline
\end{tabular}
\end{table}

In this section, we present the design for \rp, a GNN inference framework for streaming graphs that supports \textit{trigger-based} applications that need to be notified of changes to the predictions of any graph entity upon receiving updates as soon as possible. \rp uses incremental computation over a small batch of updates,
intelligently using the delta in prior embeddings
to avoid redundant computation of parts of the computation neighborhood for linear aggregation functions. \rp supports both an efficient single machine execution if the entire graph, its features, and embeddings fit in the RAM of a single server, as well as a distributed execution to scale to larger graphs with efficient communication primitives.

\subsection{Assumptions}
\label{subsec: prelim}

We make certain simplifying assumptions when designing \rp. 
The GNN model used for inferencing has $L$ layers. As bootstrap, the initial embeddings have been calculated for all existing graph entities prior to updates arriving. Each vertex $u \in V$ has its features $x_u$ and intermediate and final layer embeddings $h_u^l$. The initial embeddings $H_{T_0}^l$ of layer $l$ are generated and stored using the trained model, and the current labels of all the vertices can be extracted from $H_{T_0}^L$. These serve as starting points when new vertex/edge updates arrive.

\rp is designed for GNN models that use \textit{linear aggregation functions}. While our approach supports all linear functions, we use $sum$ to explain the propagation of updates. We use a GNN model that performs vertex classification as an example, though this can be extended to other vertex- or edge-based tasks.
We currently support three types of popular graph updates: \textit{edge additions}, \textit{edge deletions}, and \textit{vertex feature changes}. E.g. while the set of users in a fintech or social network (vertices) do not vary often, the interactions among them and their own features change regularly. 
Support for vertex additions and deletions 
is left to future work.

We assume that updates arrive \textit{continuously} and are \textit{batched} into fixed batch sizes that are then applied to the graph, triggering (incremental) computation of the affected entities. The updated predictions are immediately made available to the consumers. Since we assume a high update rate of $100$--$1000$s per second, this bulk operation can amortize the overheads, reduce redundant computation, and achieve a higher throughput. The size of the batch is a hyper-parameter that can be tuned to trade-off between update throughput that can be processed and the batch execution latency. We evaluate \rp for different batch sizes in our experiments. This approach can be extended to pick a dynamic batch size based on an elapsed time-period or latency deadlines, and is left as future work.

\begin{figure}[t]
    \centering
        \includegraphics[width=\columnwidth]{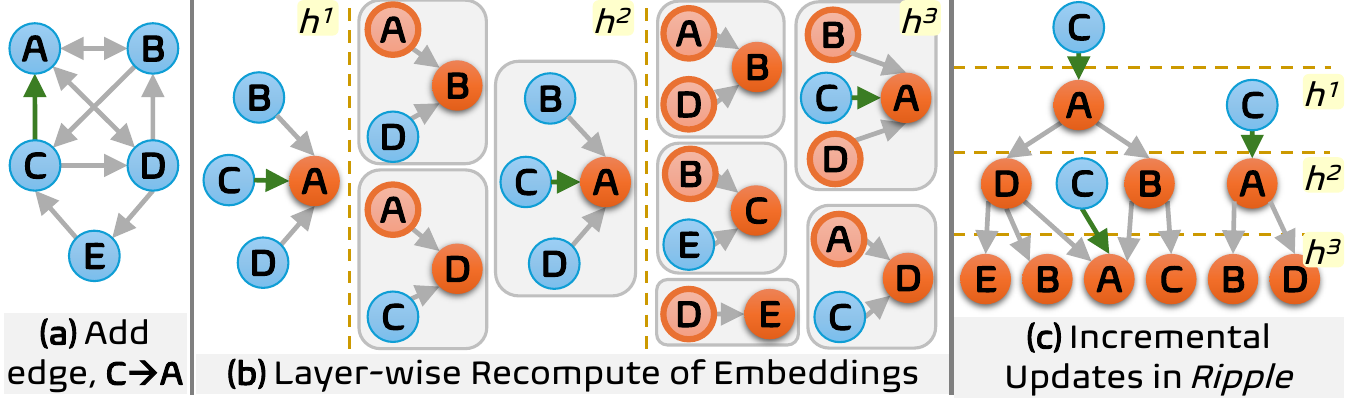}
        \caption{Contrasting \textit{(b) Recomputation (RC)} and \textit{(c)} \rp's \textit{Incremental Computation} for \textit{(a)} Edge addition in graph. Dark orange indicates vertex embeddings being updated in a hop. 
        }
        \label{fig:ripple-single-schematic}
\end{figure}

\subsection{Layer-wise Update Propagation through Recomputation}

We first describe a competitive baseline approach that performs layer-wise \textit{recomputing} scoped to the neighborhood of updates.
When an edge topology or vertex feature is updated, this causes cascading updates in the embeddings of the neighborhood of the vertex on which the update is incident, as shown in Fig.~\ref{fig:streaming-edge-add}. For instance, an \textit{edge addition} $(u, v)$ immediately alters the $h^1_v$ embedding of the sink vertex $v$, which leads to further downstream changes. This is illustrated in Fig.~\ref{fig:ripple-single-schematic}(b) where adding the edge $C \rightarrow A$ initially changes only the $h^1_A$ embeddings of sink vertex $A$ at hop distance 1 from source vertex $C$. However, recomputing $h^1_A$ requires fetching the $h_0$ embeddings of all the in-neighbors of $A$, viz., $h^0_B$, $h^0_C$, and $h^0_D$, and performing the aggregate operation on them followed by the update operation.

Similarly, for the next layer of updates to $h^2$ embeddings, the update to $h^1_A$ cascades to all out-neighbors $\{B, D\}$ for $A$, causing $h^2_B$ and $h^2_D$ to be updated by aggregating the new value of $h^1_A$ and prior value of $h^1_D$ for the neighbor $B$, and the new value of $h^1_A$ and prior value of $h^1_C$ for neighbor $D$.
In addition, the addition of the $C \rightarrow A$ edge will also cause the $h^2_A$ embeddings for $A$ to get updated, aggregating the new edge embedding from $h^1_C$ and prior values of $h^1_B$ and $h^1_D$. 
Hence, recomputation of the $h^2$ embeddings requires pulling the $h^1$ embeddings of all the in-neighbors of the affected vertices, $A, B$, and $D$. Similarly, the $h^3$ embeddings of the out-neighbors of $\{A, B, C, D, E\}$ will be updated by pulling the $h^2$ embeddings of their neighbors and recomputing the aggregation function over all of them, followed by the update function. These cascading changes are illustrated by the orange vertices in Fig.~\ref{fig:ripple-single-schematic}(b) at each layer.

An \textit{edge deletion} will also affect the same set of vertices at each hop and requires similar recomputations, except now there will be \textit{one less neighbor} when updating $h^l_v$. The embedding update to $h^1_v$ will cause a similar cascading set of changes to downstream vertices. E.g., if the edge $C \rightarrow A$ is removed at a later time, the only vertex whose $h^1$ embeddings would change is $A$, followed by a similar propagation pattern as when the edge was added.

Lastly, when a \textit{vertex feature} is updated, the scope of the out-neighborhood that is affected is different from that of an edge update. While an edge update $(u, v)$
impact only the sink vertex $v$ at hop $1$ from $u$, a vertex feature update $(u, x'_u)$ impacts all the out-neighbors of $u$ at hop distance $1$. For instance, in Fig.~\ref{fig:ripple-single-schematic}(b), after the edge addition $C \rightarrow A$ is processed, if the features of vertex $C$ change, it would impact both its out-neighbors, $A$ and $D$. This will lead to a much larger propagation neighborhood downstream. Consequently, vertex updates can be slower to apply. From an algorithmic perspective, both types of updates can be thought of as originating from vertex $u$ as the hop-$0$ or source vertex.

A key limitation of layer-wise recomputation is that updating the $h^l_v$ embedding of vertex $v$ at layer $l$ requires us to fetch the embeddings for the previous layer $h^{l-1}_u$ of \textit{all its in-neighbors} $u$, regardless of whether their $h^{l-1}$ embeddings were updated or not, as long as some of them were updated. \textit{This leads to wasted computation because updating even a single in-neighbor among many requires aggregating the embeddings of all in-neighbors to compute the new embedding for the sink vertex, and this grows exponentially downstream.} 

Next, we describe the incremental computation model of \rp that avoids this by reducing the number of operations performed during aggregation to be proportional to the number of vertices updated through a strictly \textit{look-forward} model.

\subsection{Incremental Update Propagation in \rp}
\label{sec:design:inc}
\rp treats vertices as first-class entities that are responsible for managing and updating the embeddings they own. It employs a strictly \textit{look-forward} computation model to handle update propagation. 
When building the propagation tree due to an update,
at each hop \textit{l} from the update, the affected vertices apply the \textit{messages} received from the previous hop to compute their updated embeddings. \rp does not perform pruning or selective updates, i.e., it updates all affected vertices at each hop by applying the received messages to ensure deterministic and accurate computation of embeddings. These messages carry incremental updates rather than just the raw embeddings, as we describe later. The receiving vertices then calculate and propagate messages to their own out-neighbors' \textit{mailboxes}. These out-neighbors form the next hop of the propagation tree.
This computation model leverages the linearity of underlying aggregation functions, which allows messages at a vertex to be efficiently and incrementally aggregated. 

Specifically, each vertex maintains $L$ \textit{mailboxes}, one for each hop, to accumulate messages from the impacted vertices in the previous hops. Rather than pulling raw embeddings from all in-neighbors at the previous hop, as is done in layer-wise recomputation, each impacted vertex at the current hop receives and processes only incremental messages from its impacted in-neighbors at the previous hop. It then uses its current embedding along with the cached aggregate messages in its mailbox to perform minimal incremental computation to determine its updated embedding. We detail this next.

\begin{figure}[t]
    \centering
    \includegraphics[width=0.7\linewidth, trim={0cm 15cm 15cm 2cm}, clip]{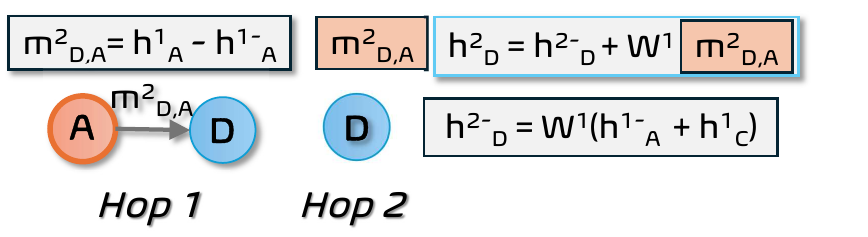}
    \caption{Using messages in the \textit{mailbox} to update the hop-2 embeddings of vertex $D$~(in orange). The effect of the old embedding of $A$ is negated and replaced with the new embedding.}
    \label{fig: incremental-update-example}
\end{figure}

\subsubsection{\rp Messages}
Propagation of updates in \rp can happen due to multiple update conditions. The purpose of a message is to \textit{nullify} the impact of the old embeddings~($h^{l-}$) and \textit{include} the contribution of the current embeddings~($h^{l}$) on the downstream vertices at hop distance $l$. 
We describe these update conditions and their consequent actions next, using a $sum$ aggregation function as an example. 

\paragraph{The embedding of vertex $u$ is updated from $h^{l-}_u$ to $h^{l}_u$} This change can occur due to a direct update to the feature of $u$, or because the vertex $u$ falls within the propagation tree of another edge or vertex update. In the case where vertex $u$ is not at the final hop of propagation of another update, vertex $v$ receives a message $m^{l + 1}_{v,u}$ in its hop ($l+1$) \textit{mailbox} from vertex $u$ because it was updated, and because there is an edge $(u, v)$ connecting the two vertices. Simply put, the contents of $m^{l + 1}_{v,u}$ is meant to invalidate the effect of the old embedding $h^{l-}_u$ on $h^{l+1}_v$ and replace its contribution with the new embedding $h^{l}_u$. 

In Fig.~\ref{fig: incremental-update-example},
for a $sum$ aggregation function, vertex $D$ receives the message $m^2_{D, A} = (h^1_A - h^{1-}_A)$. When combined with $h^{2-}_D$, it generates the new embedding $h^2_D$ for $D$.
\textit{The key insight here is that, unlike layer-wise recomputation where a single update would have caused $k$ additions to be performed over the embeddings from $k$ in-neighbors as part of aggregation, here we only perform one subtraction and one addition when applying one update for a neighbor, thus performing incremental rather than whole-neighborhood recomputation.}
In case vertex $u$ is at the final hop of the propagation tree of another update, no messages need to be sent further downstream.

\paragraph{An edge $(u, v)$ was added/deleted to/from the graph} Like the previous case, here again $m^{l + 1}_{v,u}$ will nullify the effect of $h^{l-}_u$ on $h^{l+1}_v$. But before the edge was added, vertex $h^{l+1}_v$ does not have any contribution from vertex $u$. So this is a simplified variant of the previous scenario, except that the old embedding of $u$~($h^{l-}_u$) is assumed to be zero. Similarly, deleting an edge $(u, v)$ means that $v$ must later not get any contribution from $u$. This too is a flavor of the previous case, but with the new embedding of $u$~($h^l_u$) set to be zero. 

\paragraph{Combining multiple updates at a vertex} Lastly, we discuss the case where a vertex $v$ receives messages from multiple vertices at hop $l$. Due to the inherent \textit{permutation-invariance} of the GNN aggregation functions, i.e., following a commutative property, the arriving messages can be accumulated in a vertex's mailbox in any order. \rp's \textit{propagate operation} handles this by accumulating messages from all sender vertices at hop $l$ in the receiver's mailbox. In Fig.~\ref{fig: incremental-update-example},
in addition to $m^2_{D, A}$, vertex $D$ could also receive a message from $C$, $m^2_{D, C}$ due to an update of $h^{1-}_C$ to $h^1_C$. The messages that arrive at $D$'s hop $l$ mailbox are added. The resultant $h^2_D$ is identical to the outcome if the updates had been applied individually.

This incremental computation model can easily be generalizable to other linear aggregators beyond $sum$.
For both $mean$ and $weighted~sum$ each message includes a weight $\alpha$ when propagated to a neighbor. The message to propagate a change from $u$ to $v$ at $l$ hop can then be modeled as $m^{l + 1}_{v, u} = \alpha h^{l}_u - \alpha h^{l-}_u$ to effectively replace the effect of the old embedding with the new embedding of $u$ at $v$.

\subsubsection{\rp Operations}
\rp consists of two primary operators -- \textit{update} and \textit{propagate}. The \textit{update} operator applies changes to the graph's topology or features at hop $0$. In addition, it computes and sends messages to the mailboxes of impacted vertices at hop $1$.
The \textit{propagate} operator then handles the task of propagating all relevant changes to the subsequent hops. At each hop $l$, this process involves two phases -- \textit{apply} and \textit{compute}. During the apply phase, each vertex applies the messages accumulated in its $l$-hop mailbox to get its updated embedding at hop $l$. Using this new embedding, in the \textit{compute} phase, the vertex generates and sends messages to its out-neighbors' $l+1$-hop mailboxes.
For instance, due to the update to $h^1_A$ at hop 1 in Fig.~\ref{fig: incremental-update-example}, the propagate operator computes the message for $A$'s out-neighbors $\{B, D\}$ and send the message $m^2_{B, A}$ and $m^2_{D, A}$ to their hop 2 mailboxes. During the processing of $D$ at hop $2$, it then applies the message $m^2_{D, A}$ at $D$, to $h^{2-}_D$, to get the new embedding $h^2_D$ as shown.

\subsubsection{Benefits Analysis}

In layer-wise recompute (RC), for each update to a vertex $u$ at hop $l+1$, the aggregation function, $\oplus_{v \in N_{in}(u)} (h^{l}, v)$ is applied to all its in-neighbors $N_{in}(u)$. This performs $k$ numerical operations, where $k$ is the vertex's in-degree. In contrast, for \rp we have $\oplus_{v \in N'_{in}(u)} (\ominus(h^{l-}, h^{l}, v), v)$. Here, $\ominus$ captures the delta change by negating the impact of the previous embedding, and $\oplus$ aggregating the delta.
E.g., for \textit{sum}, $\ominus$ is just a subtraction operation and $\oplus$ an addition operation, while for \textit{mean} and \textit{weighted sum}, as discussed above, these are weighted negation and weighted sum operations.
These are each performed $k'$ times to give $2k'$ numerical operations, where $k'=|N'_{in}(u)|$, the number of \textit{updated} vertices in the in-neighborhood of $u$. We typically have $k'\ll k$ since $N'_{in}(u) \subset N_{in}(u)$, and only a smaller subset of neighbors tend to get updated within a batch.
These benefits of \rp over RC accumulate and grow layer-wise through the update propagation tree.

\begin{figure}[t]
    \centering
    \includegraphics[width=0.7\columnwidth]{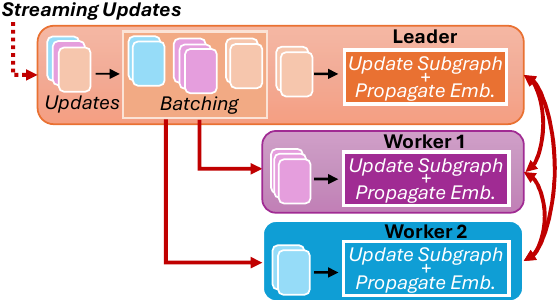}
    \caption{Distributed architecture of \rp}
    \label{fig:arch}
\end{figure}

\begin{figure}[t]
    \centering
        \includegraphics[width=\columnwidth]{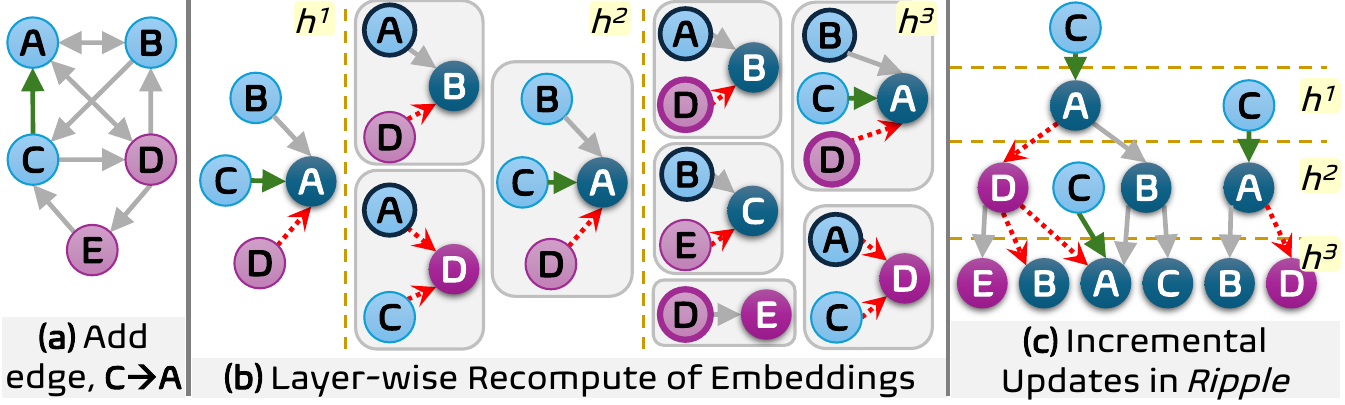}
        \caption{A distributed execution version of Fig.~\ref{fig:ripple-single-schematic}. $\{A,B,C\}$ are in one machine while $\{D, E\}$ are in another. Red dotted edges indicate remote message passing across machines.
        }
        \label{fig:ripple-distr-schematic}
\end{figure}

\section{Distributed GNN Inferencing in \rp}
\label{sec:design:distr}
Real-world graphs can be large in size. Besides the graph structure and features, we also need to store the various embeddings for all vertices to help with the incremental computation. These data structures can be resource-intensive, exceeding the RAM of a single machine. We design a distributed execution model for \rp to perform incremental computation over updates received for the graph, which is partitioned across different machines in a cluster, while retaining an execution flow similar to the single-machine inferencing (Fig,~\ref{fig:arch}).

\subsection{Partitioning the Graph}
The initial graph structure before updates start streaming and their embeddings computed for the GNN model are first partitioned and kept in-memory, across multiple machines~(\textit{workers}). We use the METIS~\cite{karypis1997metis} to partition the vertices of the graph such that the vertex count is balanced across workers and the edge cuts across partitions are minimized. This reduces network communications. Each worker is only responsible for maintaining and updating the embeddings for its local vertices. However, to reduce communication during the BFS propagation, we also replicate the boundary vertices that have edge cuts. These are called \textit{halo} vertices in DistDGL~\cite{zheng2020distdgl}. While the embeddings of the halo vertices are present on their parent worker and not the remote workers, we maintain a mailbox for these halo vertices on the remote workers as a stub.

\subsection{Request Batching and Routing}
One of the workers acts as the \textit{leader} and is responsible for receiving the stream of graph updates.
The leader creates batches of updates destined for each worker, including itself. An update request $r$ is assigned to the batch destined for worker $w_i$ if the hop-$0$ vertex~(source vertex, for an edge update, or updated vertex, for a vertex feature update) belongs to $w_i$. Since edge additions or deletions can span workers, we discuss the different possible cases. 
\textit{(1) An update request is completely local to a worker} if both the incident vertices of an edge update are local to the worker. Here, the update is assigned to that worker. A vertex update is always local to the parent worker.
\textit{(2) Edge request that span workers} have source and sink vertices  in different partitions. We assign the update to the worker that owns the hop-$0$ or source vertex of the edge update. Further, we also create a \textit{no-compute} request and send it to the partition where the sink vertex for this edge resides. This does not propagate any change but is used to update the local topology in the other partition. 
Once enough streaming updates for a batch have arrived, they are sent to the respective workers.

\subsection{Update Processing} 
Upon receiving a batch of updates, a worker applies the topology changes to the local partition. It also sends messages to the mailboxes for the hop $1$ vertices. 
For each subsequent hop $l \in [1, L]$, it is possible that the set of impacted vertices includes halo vertices. A worker $w_i$ only has access to the embeddings for the vertices local to it, and cannot access or update the embeddings of the halo vertices. Each worker puts the messages for the halo vertices into mailboxes present in their parent worker. A worker can receive messages for its local vertices from its corresponding halo vertices present in other workers; these received messages are then added to the local mailboxes of the vertices.

After receiving the messages for the hop $l$, the messages in the mailboxes are then applied, similar to a single machine scenario. Then, messages for hop $l+1$ are calculated and put in the respective mailboxes. This process repeats $L$ times to get the final layer embeddings for the impacted vertices at the $L$ hop due to the requests in the current mini-batch. 

This communication pattern resembles a Bulk Synchronous Parallel~(BSP) model, which alternates between phases of computation and communication, followed by a synchronization barrier. This is crucial because the embeddings for hop $l + 1$ can only be updated after the $l$-hop embeddings have been computed. 
\rp leverages MPI as the communication stack to facilitate efficient message passing between workers.

\section{\rp Implementation and Baselines}
\label{sec:impl}
\label{subsubsec: eval-baselines}

\rp is implemented natively in Python for both single-machine and distributed setups.
Both leverage NumPy v$2.0$. The distributed setup also utilizes MPI for inter-process communication. We only support CPU-based execution due to its competitive performance, as discussed later. 

We also implement the \textit{vertex-wise} and \textit{layer-wise recompute} inference strategies as baselines using the state-of-the-art (SOTA) DGL~\cite{wang2019deep} GNN training and inference framework. 
We use DGL v$1.9$ with a PyTorch backend.
Additionally, we also use our own implementation of the \textit{layer-wise recompute} strategy, because of the high overheads we observe for graph updates using DGL~(Fig.~\ref{fig: gpu-analysis}). 
While InkStream~\cite{wu2023inkstream} also explores GNN inferencing on streaming graphs, it is limited to \textit{max/min} aggregation functions, and therefore, cannot be used as a baseline for popular GNNs with linear functions.

To keep the comparison fair, all strategies only update the embeddings of impacted vertices at the final hop when performing a batch of updates. We consider the whole neighborhood of a vertex at each hop during inferencing, without sampling or any approximations, to ensure that the resulting predictions are accurate and deterministic. Finally, \rp calculates \textit{accurate} embeddings at all hops within the limits of floating-point precision.

\section{Evaluation}
\label{sec:evaluation}
We evaluate \rp on diverse graph datasets and GNN workloads for single machine and distributed setups and compare it against contemporary baselines.

\subsection{Experimental Setup}
\label{subsec: eval-expsetup}

\begin{table}[t]
\centering
\setlength{\tabcolsep}{2pt}
\caption{Graph datasets used in experiments.} 
\begin{tabular}{l||r|r|r|r|r}
\hline
\textbf{Graph} & $|V|$  & $|E|$   & \# Feats. & \# Classes & Avg. In-Deg.\\ \hline\hline
\textbf{\textit{Arxiv}}~\cite{hu2020open}                & $169K$   & $1.2M$  & $128$ & $40$ & $6.9$\\
\textbf{\textit{Reddit}}~\cite{hamilton2017sage}               & $233K$   &  $114.9M$  & $602$ & $41$ & $492$\\
\textbf{\textit{Products}}~\cite{hu2020open}         & $2.5M$   & $123.7M$ & $100$  & $47$ & $50.5$\\ 
\textbf{\textit{Papers}}~\cite{hu2020open}          & $111M$   & $1.62B$ & $128$  & $172$ & $14.5$\\  \hline
\end{tabular}
\label{tab:dataset-specs}
\end{table}

\subsubsection{GNN Workloads}
\label{subsubsec: eval-workloads}
We use three popular GNN models for vertex classification: \textit{GraphConv}~\cite{kipf2016semisupervised}, \textit{GraphSAGE}~\cite{hamilton2017sage} and \textit{GINConv}~\cite{xu2018powerful}. Each of these is paired with common linear aggregation functions like sum, mean, and weighted sum to give $5$ representative workloads: 
\underline{G}raph\underline{C}onv with \underline{S}um~(\textbf{GC-S}), 
\underline{G}raph\underline{S}AGE with \underline{S}um~(\textbf{GS-S}), 
\underline{G}raph\underline{C}onv with \underline{M}ean~(\textbf{GC-M}),
\underline{G}IN\underline{C}onv with \underline{S}um~(\textbf{GI-S}), and 
\underline{G}raph\underline{C}onv with \underline{W}eighted Sum~(\textbf{GC-W}). 
We train these models on a snapshot of the graph datasets listed below, using $90\%$ of all the edges selected at random, and extract the embeddings, which make the initial state for inferencing.

\begin{figure}[t]
    \centering
    \subfloat[Arxiv\label{subfig: gpu-analysis-arxiv}]{\includegraphics[width=.28\linewidth]{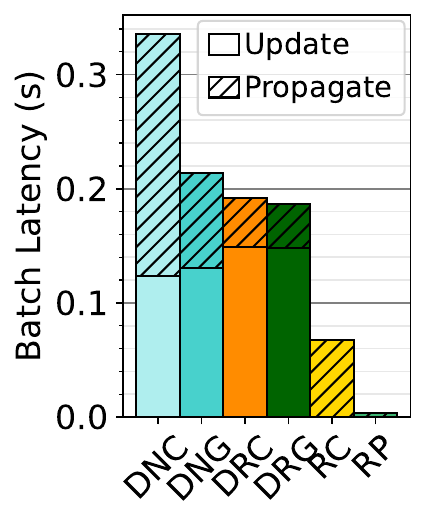}}\qquad
    \subfloat[Products\label{subfig: gpu-analysis-products}]{\includegraphics[width=.28\linewidth]{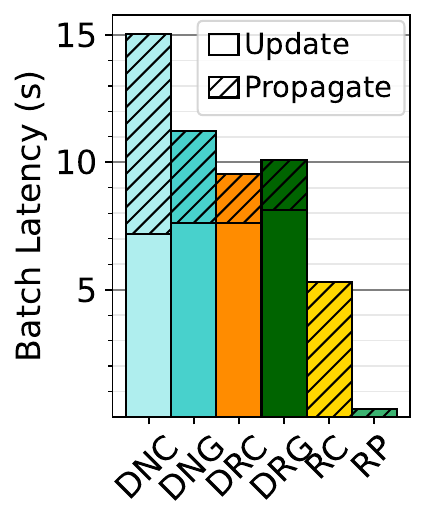}}
    \caption{Comparison of DGL's vertex-wise and layer-wise recompute, our layer-wise recompute and \textit{Ripple}'s incremental compute on CPU and CPU+GPU for \textit{GC-S} with $3$ layers and batch size $10$ for Arxiv and Products graph.
    }
    \label{fig: gpu-analysis}
\end{figure}

\subsubsection{Datasets}
\label{subsubsec: eval-datasets}
We use $3$ well-known graph datasets for our single machine evaluation: ogbn-arxiv~(\textit{Arxiv})~\cite{hu2020open} which is a citation network, \textit{Reddit}~\cite{hamilton2017sage} which is a social network, and ogbn-products~(\textit{Products})~\cite{hu2020open} which is an eCommerce network~(Table~\ref{tab:dataset-specs}).
We remove a random $10\%$ of edges from these graphs; the graph with all vertices and the remaining 90\% of edges serves as the initial snapshot to which streaming updates arrive. The 10\% of removed edges are then streamed as edge additions during the evaluation. We also pick a random set of edges from the snapshot that we delete and a random subset of vertices whose features we update. We generate $90K$ updates for each graph, with equal numbers of edge additions, edge deletions, and vertex feature updates in random order. 

We use a fourth large-scale citation network, ogbn-papers100M~(\textit{Papers})~\cite{hu2020open}, that does not fit on RAM of a single machine. This is studied for the distributed execution and scaling. Here, we use $50\%$ of the graph edges as our initial snapshot but otherwise, prepare the updates to be streamed in a similar fashion as above.

\begin{figure*}[t]
    \centering
    \includegraphics[width=\linewidth]{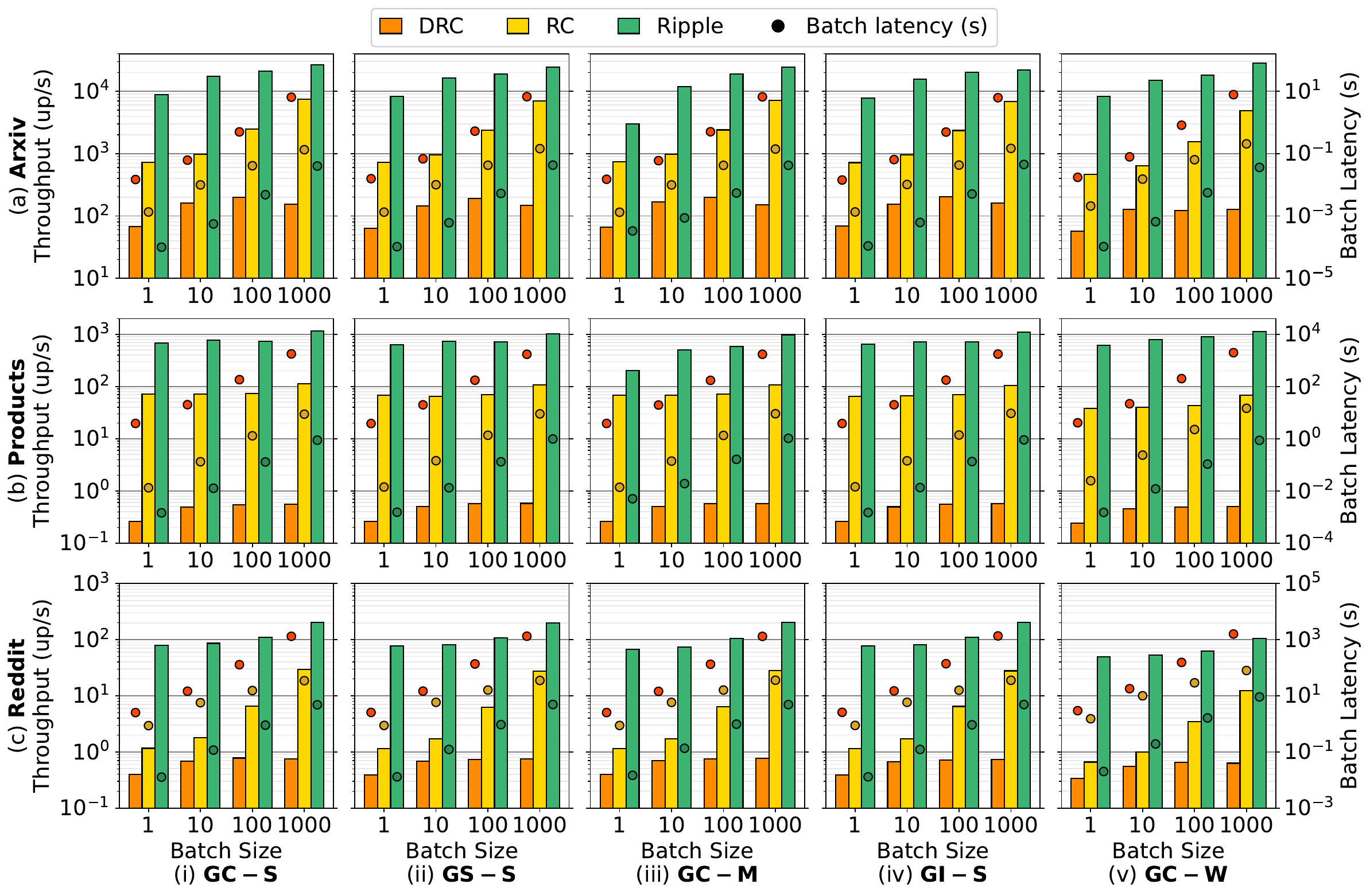}
    \caption{Single machine performance of \rp against \textbf{DRC} and \textbf{RC} on the 2-layer variants of $5$ GNN workloads for Arxiv, Reddit and Products across batch sizes. Left y-axis shows \textit{Throughput} (updates/second, bar, log scale) while right y-axis has \textit{median batch latency} (seconds, marker, log scale)}
    \label{fig:tp-lat-l2}
\end{figure*}

We perform GNN inferencing over the initial graph snapshot to generate the intermediate and final layer embeddings for all the vertices; the snapshot and embeddings are kept in memory. In the distributed setup, we partition the graph using METIS into the required number of parts and load the local subgraph, their embeddings, and the halo vertices in memory.

\subsubsection{Hardware Setup}
\label{subsubsec: eval-testbed}
All single-machine experiments are conducted on a server with a $16$-core Intel Xeon Gold CPU~($2.9$GHz) with $128$GiB of RAM. 
The distributed execution is performed on a cluster of $16$ such compute servers, connected by a 10~Gbps Ethernet.

We also compare the DGL implementation of vertex-wise and layer-wise recompute using both CPU and GPU to demonstrate that the GPU-based execution is occasionally slower than the CPU-based execution, and that both are slower than our custom implementation of layer-wise recompute on the CPU. For this, we use a GPU workstation with a $12$-core AMD Ryzen $9$ $7900$X CPU~($4.7$GHz) with $128$GiB RAM, having an NVIDIA RTX $4090$ GPU card with $24$GiB GPU memory.

\begin{figure*}[t]
    \centering
    \includegraphics[width=\linewidth]{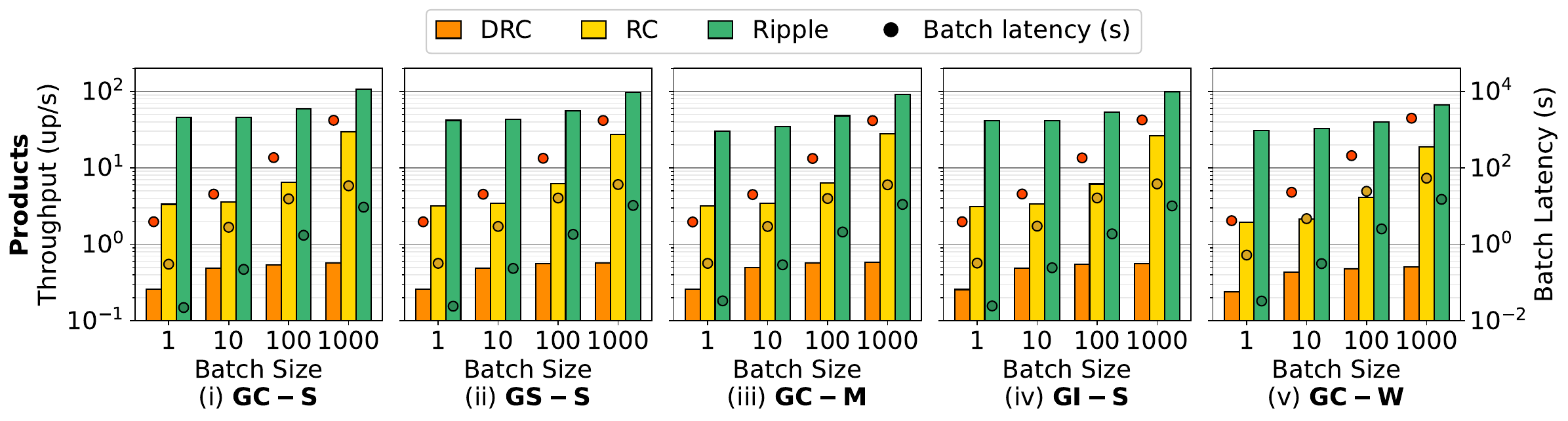}
    \caption{Single machine comparative performance 
    for 3-layer GNN workloads for Products.
    }
    \label{fig:tp-lat-l3}
\end{figure*}

\subsection{Comparison with SOTA Frameworks on CPU and GPU}
\label{subsec: eval-comparison_with_sota}
We first compare DGL's vertex-wise inference on the CPU (\textbf{DNC}) and CPU+GPU (\textbf{DNG}), with its layer-wise recompute strategy on the CPU (\textbf{DRC}) and CPU+GPU (\textbf{DRG}).
Further, we also contrast these against our implementation of layer-wise recompute (\textbf{RC}) and the \rp incremental strategies (\textbf{RP}) run on the CPU. All of these execute on the GPU workstation using a single-machine execution model.

Fig.~\ref{fig: gpu-analysis} shows the median batch latency of processing $100$ batches of $10$ updates each by these strategies for the Arxiv and Products graphs. 
The vertex-wise inference strategies DNC and DNG, compared against their corresponding layer-wise recompute strategies DRC and DRG are significantly slower~($\approx74\%$ and $\approx58\%$ on the CPU, $\approx17\%$ and $\approx12\%$ on the CPU+GPU, for Arxiv and Products, respectively). 
This is because of the redundant computations that are performed for each batch in vertex-wise inference. We also notice that the GPU-based DRG method offers limited benefits over the CPU-based DRC strategy.
Since the layer-wise strategy processes embeddings one layer at a time, it only requires data for the immediate previous layer in the computation graph. 
Due to these small batch sizes, the computational workload is minimal, which limits the performance gains from GPU-based computation~($\approx5\%$ faster for Arxiv, but $\approx6\%$ slower for Products). This justifies our choice of a layer-wise approach and just using CPUs.

Further, our own custom layer-wise recompute implementation on CPU (RC) is $\approx40$--$60\%$ faster than both the CPU and GPU versions of DGL's layer-wise recompute~(DRC, DRG). 
While DGL's graph APIs ease the development of GNN training models, they are not optimized for handling a stream of updates.
As a result, updating the graph topology consumes a significant amount of time~(\textit{Update} stack in Fig.~\ref{fig: gpu-analysis}). In contrast, our RC implementation uses lightweight edge list structures designed to efficiently handle streaming updates. They offer much faster update times and comparable or slightly slower compute propagation times.
As we discuss later, the incremental computation of \rp (RP) is substantially faster than all of these.
In the rest of the experiments, we use DRC and RC running on the CPU as the competitive baselines to compare against \rp's incremental approach.

\begin{figure}[t]
    \centering
    \subfloat[Products with GC-S 2-layers\label{subfig: nodes-v-latency-l2}]{\includegraphics[width=.36\linewidth]{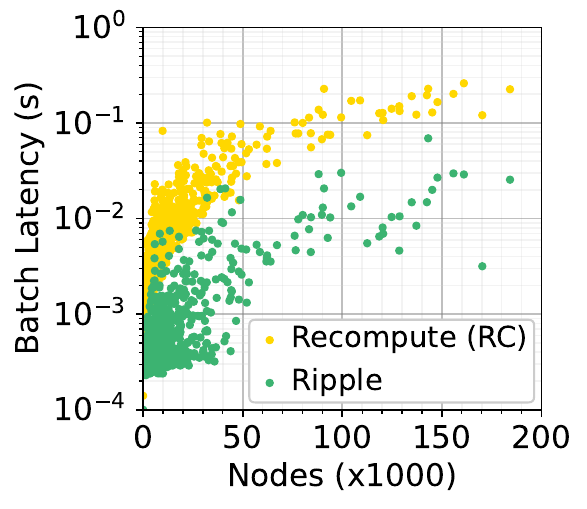}}\qquad
    \subfloat[Products with GC-S 3-layers\label{subfig: nodes-v-latency-l3}]{\includegraphics[width=.36\linewidth]{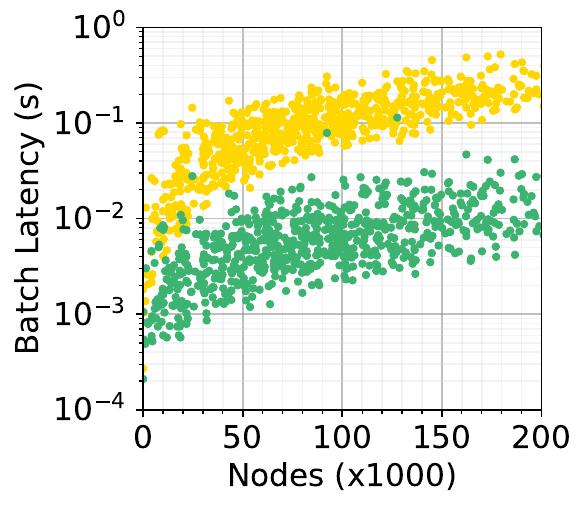}}
    \caption{Effect of \# of vertices in the propagation tree on the batch latency for Products using GC-S with a batch size of 1}
    \label{fig: nodes-v-latency}
\end{figure}

\begin{figure}[t]
    \centering
    \subfloat[\textit{Throughput} (left Y axis, bar) and \textit{latency} (right, marker) 
    of \rp and RC on 8 partitions for  $3$-layer workloads\label{subfig: tp-lat-l3-papers8}
    ]{
        \includegraphics[width=0.7\linewidth]{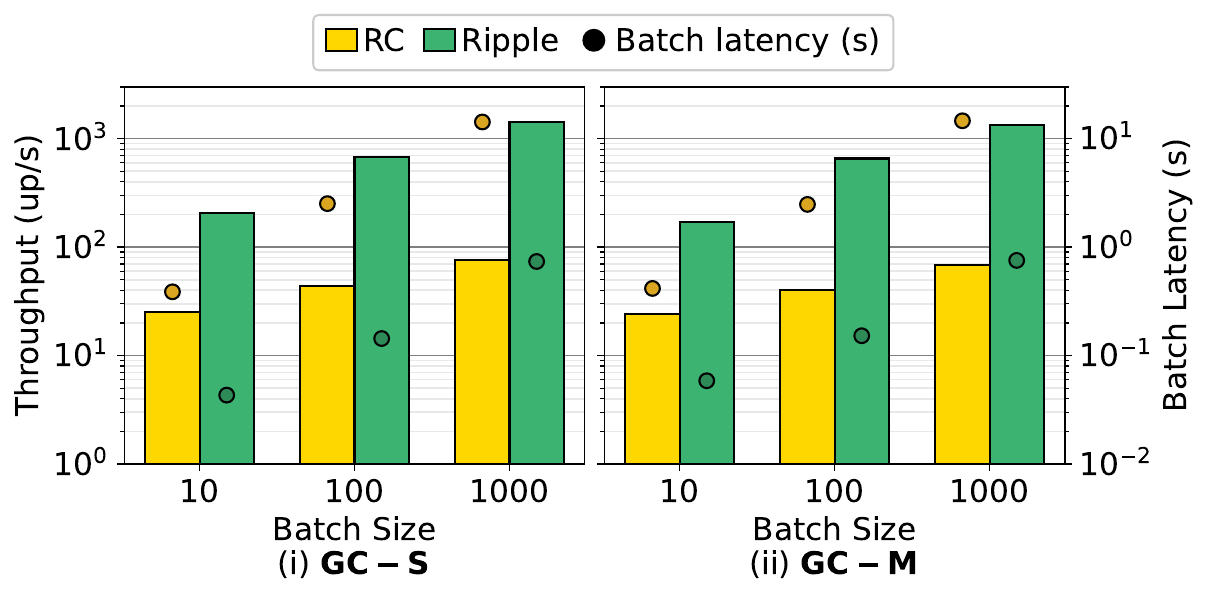}
    }\\
    \subfloat[Strong scaling of GC-S-3L with \# of partitions, for $3$ batch sizes.\label{subfig: scale}]{
        \includegraphics[width=.37\linewidth]{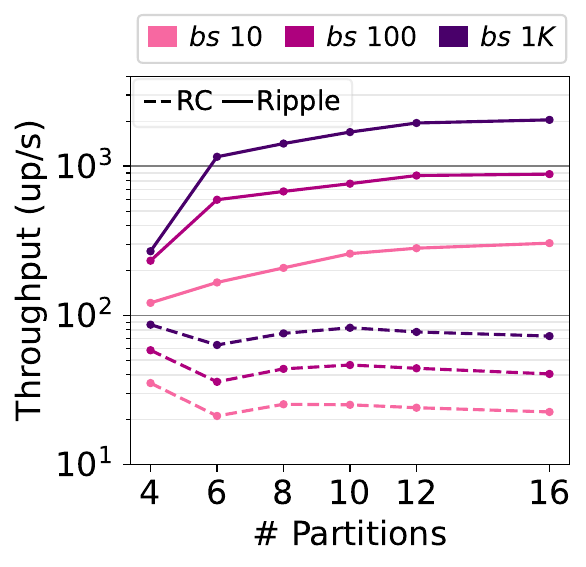}
    }
    ~
    \subfloat[Compute and communication times with \# of parts, for GC-S-3L, bs=$1k$\label{subfig: comp-comm-bkdown}]{\includegraphics[width=.4\linewidth]{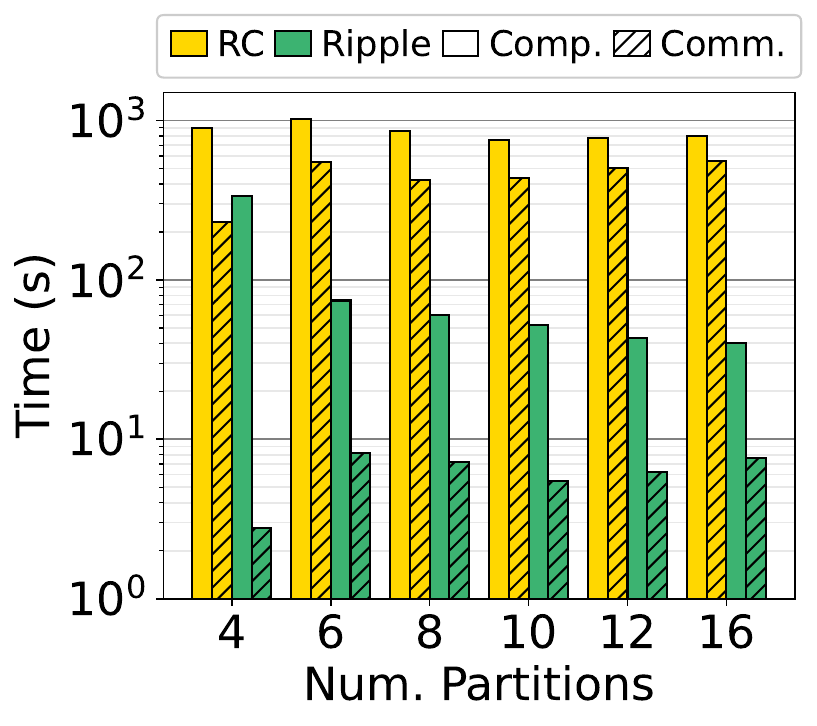}}
    \caption{Distributed scaling performance of \rp against RC on Papers dataset.}
    \label{fig: dist-plots-papers}
    \end{figure}

\subsection{Single Machine Performance}
Figs.~\ref{fig:tp-lat-l2} and~\ref{fig:tp-lat-l3} report a comprehensive performance comparison of \rp against the DRC and RC baselines, for all $5$ workloads with 2-layers, using $4$ batch sizes, $\{1,10,100,1000\}$.
For brevity, we limit reporting results for 3-layer GNN models to just Products, in Fig.~\ref{fig:tp-lat-l3}. 

We see an increase in average throughput across batch sizes for Arxiv, Products, and Reddit graphs by up to $150\times$, $2000\times$ and $190\times$ for \rp over DRC, and up to $14\times$, $19\times$, and $38\times$ for \rp over RC for the 2-layer workloads~(Fig.~\ref{fig:tp-lat-l2}). These reflect the benefits of incremental computing of \rp.
\rp can achieve throughput as high as $\approx1200$up/s for Products and $\approx28000$up/s for Arxiv; it is somewhat lower for Reddit at $\approx210$up/s due to its high in-degree ($\approx500$), which has high computational cost due to the large neighborhood size.
\rp also has a competitive median batch processing latency across batch sizes, ranging between $\approx0.1$--$40$ms for Arxiv, $1.5$ms--$1$s for Products, and $\approx12$ms--$10$s for Reddit. This again correlates with the in-degree of the graphs. These latencies for \rp are small enough to be used for near-realtime applications. As expected, throughput and latency offer a trade-off as batch sizes increase.

For Products, using 3-layer GNN workloads~(Fig.~\ref{fig:tp-lat-l3}), the average throughput for \rp improves by up to $\approx140\times$ and $11\times$ against DRC and RC. This is because \rp avoids fetching and processing all in-neighbors at each hop, and this is amplified as the number of layers increases. 
This is clearly seen in Fig.~\ref{fig: nodes-v-latency}, where we compare the batch latency for RC and \rp (Y-axis) as the total number of affected vertices in the propagation tree increases~(X-axis). There is a strong correlation between these two metrics, and critically, \rp is an order of magnitude faster on latency than RC across the spectrum of vertex update ranges for both 2 and 3-layer GC-S on Products.

However, in a worst-case scenario where updates propagate to an extent where all vertices in the graph appear at a hop in the propagation tree, \rp can degrade to RC for that and further hops.  In addition, we notice in both Fig.~\ref{fig:tp-lat-l2} and~\ref{fig:tp-lat-l3} that the throughput of DRC stabilizes at a batch size of $10$ for all workloads and does not offer benefits beyond this. This is because the batching benefits are offset by the overheads of performing a large number of graph updates in a larger batch. 
The benefits of \rp do come at the cost of a higher memory usage due to maintaining additional data structures for all $L$-layer embeddings and the mailboxes for updated vertices, e.g., having an overhead of $4$GB over RC for \textit{Products} for $3$-layer GC-S. When the memory is a constraint, we can use the distributed setup,
although at a cost of performance.

\subsection{Distributed Setup Performance}
\label{subsec: eval-dist}
Papers takes $\approx500$GiB of memory to store the graph structure and all embeddings for its $111M$ vertices, This does not fit in the $128$GiB RAM available in our single server and requires distributed execution.  
Fig.~\ref{fig: dist-plots-papers} 
compares the performance of distributed \rp against a distributed version of RC we implement for Papers; Distributed DGL (DistDGL)~\cite{zheng2020distdgl} does not support online graph updates and hence we omit the DRC baseline.
The relative throughput improvement of distributed \rp over RC in Fig.~\ref{subfig: tp-lat-l3-papers8} for GC-S and GC-M workloads is similar, with a peak of $\approx30\times$ for a batch size of $1000$ for $16$ partitions~(Fig.~\ref{subfig: scale}).
This is expected given that \rp not only reduces recomputations, it also spends $\approx70\times$ less time in the communication phase~(Fig.~\ref{subfig: comp-comm-bkdown}). 
This is because RC can pull embeddings from in-neighbors in another partition that were unaffected at the previous hop but used during recomputing, while \rp avoids this.
The distributed execution also suffers from a slightly higher initial latency for transferring updates from the leader to the workers over the network. 

We also perform a strong scaling study for both RC and \rp across different numbers of partitions~($4$--$16$), reported in Figs.~\ref{subfig: scale} and \ref{subfig: comp-comm-bkdown}. We see that \rp scales well, with the throughput for $16$ being $8\times$ better than with $4$ partitions. But RC spends a major chunk of its time communicating a large number of unnecessary vertex embeddings and, therefore, does not scale. 
This is evident from Fig.~\ref{subfig: comp-comm-bkdown}, where we see that neither the computation nor the communication stacks reduce significantly for RC with the number of partitions. However, for \rp, we notice a gradual decrease in the computation bar as the partitions increase. As expected, the communication time increases gradually with the number of partitions since the number of edge cuts also increases. While the compute time bar in \rp is solely dependent on its local vertices, for RC, the compute depends on the number of in-neighbors of the vertices currently being processed.

\begin{figure}[t]
    \centering
    \subfloat[
    \textit{Throughput} (left Y axis, bar) and \textit{latency} (right, marker) of \rp and RC on 8 partitions\label{subfig: tp-lat-l3-products}]    {\includegraphics[width=.44\linewidth]{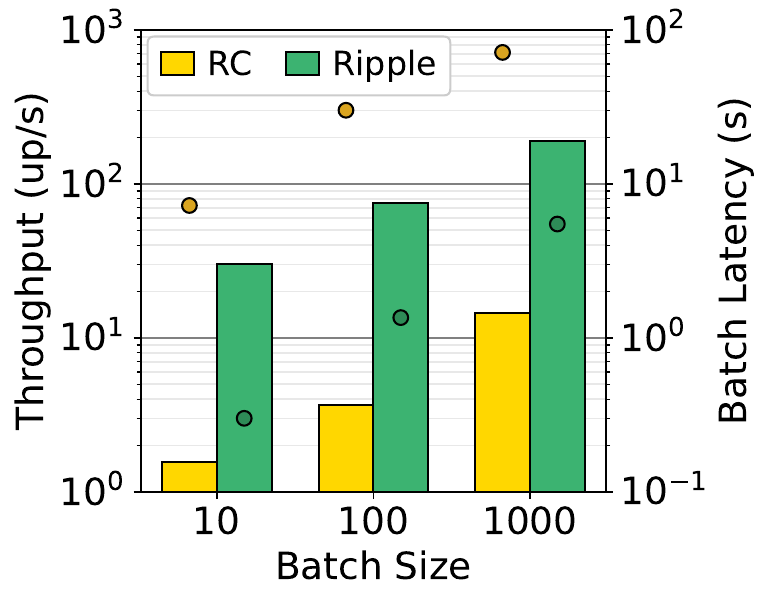}}~
    \qquad   
    \subfloat[
    Strong scaling of compute and communication times with \# of parts bs=$1k$
    \label{subfig: scale-products}]{\includegraphics[width=.345\linewidth]{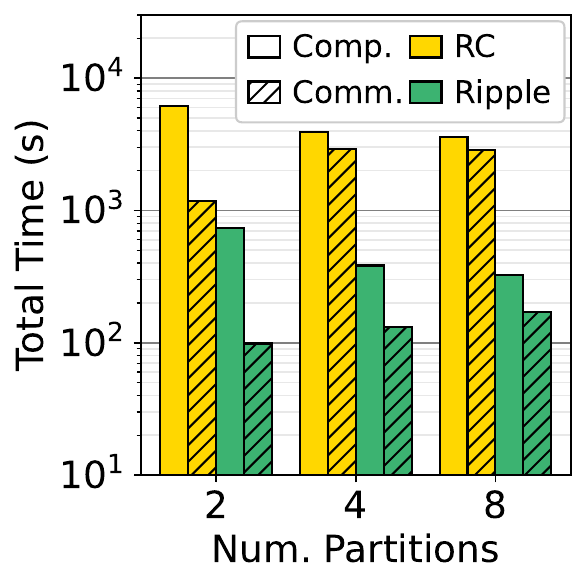}}
    \caption{Distributed scaling of GC-S-3L on Products.
    }
    \label{fig: dist-plots-products}
\end{figure}

Lastly, we study the performance of distributed \rp on Arxiv and Products with $2$, $4$, and $8$ partitions. For brevity, we plot the metrics for only Products in Fig.~\ref{fig: dist-plots-products}. While \rp does outperform RC and offers some scaling benefits for Products, it is not as pronounced as for Papers, achieving a throughput of $\approx190$up/s with 8 partitions compared to $\approx110$up/s for 2 partitions for batch size $1000$.
In fact, the corresponding single machine throughput was $\approx108$up/s, indicating that graphs that can be incrementally inferenced on a single machine, should.

\section{Conclusion and Future Work}
\label{sec: conclusion}
In this work, we present a novel framework, \rp, to efficiently perform GNN inferencing over large-scale streaming graphs. Unlike traditional methods that rely on exhaustive look-back computations, our approach adopts a strictly look-forward incremental computation, where vertices are first-class entities that manage their data and updates. \rp is able to process $\approx28000$~up/s for sparser graphs like Arxiv and up to $1200$~up/s for larger and denser graphs like Products.

We plan to extend \rp to support more update types such as vertex addition/deletion, explore streaming partitioners to maintain balanced partitions as the graph changes, and support dynamic batch sizes for latency-sensitive tasks. We intend to test \rp on larger real-world graphs like Twitter~\cite{kwak2010twitter} and compare it with other baselines such as DZiG~\cite{mariappan2021dzig}. Finally, we also plan to explore non-linear aggregation functions like max/min, along with other popular models like GAT~\cite{velivckovic2017graph}~\footnote{Acknowledgments: The authors thank Roopkatha Banerjee, Tejus C., Prashanthi S.K., and other members of the DREAM:Lab, Indian Institute of Science, for their assistance and insightful feedback on this work}.

\bibliographystyle{plain}
\bibliography{arxiv-refs}
\end{document}